\documentclass[runningheads]{llncs}
\usepackage[hidelinks]{hyperref}
\usepackage{amsmath}
\usepackage{amssymb}
\usepackage{graphicx}
\begin{document}
\title{Deep Learning based Super-Resolution for Medical Volume Visualization with Direct Volume Rendering}
\titlerunning{Super-resolution for volume rendering}
%
\author{Sudarshan Devkota
\inst{1}
\email{sudarshan.devkota93@knights.ucf.edu}\\ 
\and
Sumanta Pattanaik
\inst{1}
\email{sumant@cs.ucf.edu}}
\authorrunning{S. Devkota et al.}
%
%
\institute{University of Central Florida, Orlando, FL, USA}
\maketitle              
\begin{figure*}
  \centering
  \includegraphics[width=0.85\linewidth]{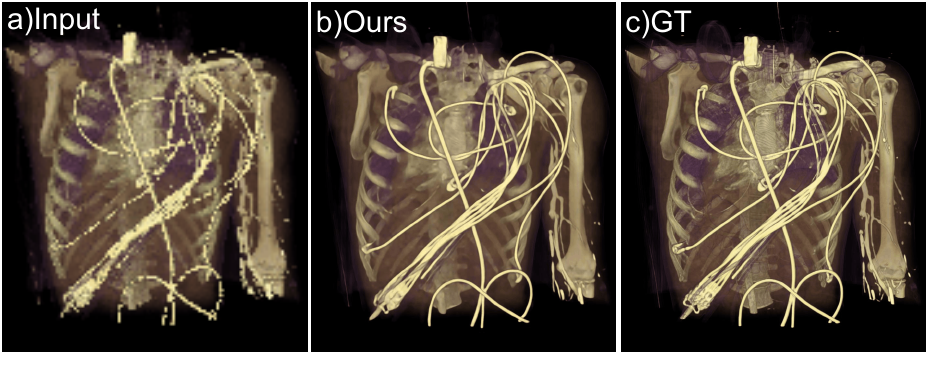}
  \centering
  \caption{Results of our super-resolution network for volumetric rendering with a) input rendering at a low resolution of 240x240 which is upscaled by a factor of 8x8 to obtain the high-resolution output b) at 1920x1920. c) is ground truth image.   }
  \label{fig:teaser}

\end{figure*}

\begin{abstract}
Modern-day display systems demand high-quality rendering. However, rendering at higher resolution requires a large number of data samples and is computationally expensive. Recent advances in deep learning-based image and video super-resolution techniques motivate us to investigate such networks for high fidelity upscaling of frames rendered at a lower resolution to a higher resolution. While our work focuses on super-resolution of medical volume visualization performed with direct volume rendering, it is also applicable for volume visualization with other rendering techniques. We propose a learning-based technique where our proposed system uses color information along with other supplementary features gathered from our volume renderer to learn efficient upscaling of a low resolution rendering to a higher resolution space. Furthermore, to improve temporal stability, we also implement the temporal reprojection technique for accumulating history samples in volumetric rendering. 
Our method allows high-quality reconstruction of images from highly aliased input as shown in figure \ref{fig:teaser}. 

\end{abstract}

\section{Introduction}

With recent advancements in imaging technology, medical volume data, such as computed tomography (CT) scans and Magnetic Resonance Imaging (MRI) images, are readily available. The rendering performed with these 3D data for visualization of anatomical structures plays a significant role in today’s clinical applications. The quality of the 3D volume data, as well as the visual fidelity of the rendered content, directly affects the diagnosis accuracy in clinical medicine. For larger volume data, the traversal of the volume becomes increasingly costly and can negatively affect the frame rate for high resolution rendering. 

In recent years, several works have addressed the goal of resolution augmentation in the medical imaging sector as a software based post-processing technique rather than an engineering-hardware issue. Such software based techniques have a variety of use cases. For instance, in cases of remote visualizations, high-resolution rendering from supercomputers can only be saved or streamed at a compressed lower resolution state due to storage and bandwidth limitations. This data, when streamed to the client-side, needs to be decompressed and upscaled in such a way that the reconstruction error is kept as low as possible. Moreover, high-resolution displays in modern-day mobile and Virtual Reality (VR) systems demand high-resolution and high-quality rendering. 


A variety of high quality image reconstruction techniques have been proposed to address this issue. Recent works in deep learning have demonstrated that learning-based image and video super-resolution methods can efficiently upscale inputs to a higher resolution when the network is trained on low and high-resolution pairs of images \cite{Dong15}. 
In image and video super-resolution literature, super-resolution is generally studied as a deblurring problem. However, unlike photographic images, each pixel sample in a rendering is a point sample in space and time which makes the final rendering to have aliasing artifacts typically at lower resolution. Thus, upscaling rendered content is considered as an anti-aliasing and interpolation problem \cite{Metapaper}. 

In our work, we investigate a deep learning based super-resolution approach for direct volume rendering (DVR) of 3D medical data. Leveraging prior works on image and video super-resolution architectures, we present a rendering pipeline that includes an artificial neural network to perform upscaling of a ray-casted visualization of medical volumetric data. Motivated by a recent work on super-sampling of surface-only rendered content \cite{Metapaper}, we plan to use the neural super-sampling architecture as a basis and extend it for volumetric rendering. Our proposed pipeline consists of a volume renderer that outputs a low-resolution rendering of medical volume data along with a number of supplementary features which enables the super-resolution network to make sensible interpretations of these features for generating a high-resolution representation of the input. Furthermore, in order to improve the temporal stability and to aid in information refill, we implement a simple, yet effective way to perform temporal reprojection for volumetric cases. This allows our network to effectively propagate and aggregate samples from neighboring frames to the current frame.

We summarize our technical contributions as follows:
\begin{itemize}
    \item We demonstrate a learning-based technique that performs up to 8x8 upsampling of highly aliased volumetric rendering with improved visual fidelity and temporal stability.
    \item We experimentally verify the effectiveness of supplementing the network with additional features to improve the quality of reconstructed image. 
    \item We implement an effective temporal reprojection technique for the accumulation of history samples in volumetric rendering.
\end{itemize}

\section{Related Work}
\label{section:relatedWork}


\subsection{Image and Video Super-resolution}
\label{section:rw-imagesr}

Deep learning-based super-resolution techniques started to gain popularity since the initial works by \cite{Dong15} where they used deep convolutional neural networks (CNN) to learn end-to-end mapping between low/high-resolution images. Several other CNN-based models have been proposed since then to improve upon the network architecture. Instead of learning the direct mapping between input and output, Kim \emph{et al.}\cite{Kim16} proposed to learn the residual between the two images by introducing a very deep network. After the introduction of residual network \cite{He16}, Zhang \emph{et al.}\cite{Zhang18a} and Lim \emph{et al.}\cite{Lim17} applied residual blocks to further improve the performance of the network. To improve upon the perceptual quality of the reconstructed photo-realistic images, Ledig \emph{et al.}\cite{SRGAN} incorporated generative adversarial networks\cite{Goodfellow14} and proposed to use a combination of loss functions including perceptual loss \cite{Johnson16} and adversarial loss\cite{Goodfellow14}. 


Video super-resolution (VSR) is more challenging compared to single image super-resolution in that one needs to gather auxiliary information across misaligned neighboring frames in a video sequence for restoration. In some recent works, recurrent networks have been widely used in video super-resolution architectures \cite{Isobe20a}\cite{Basicvsr} which naturally allows for gathering information across multiple frames. Another group of networks uses motion estimation between frames to fuse multiframe information and to improve temporal coherence. Jo \emph{et al.} \cite{Jo18} proposed to use dynamic upsampling filters for implicit motion compensation while Kim \emph{et al.} \cite{KimTH18} used a spatio-temporal transformer network for multiple frame motion estimation and warping.

\subsection{Resolution Enhancement for Rendered Content}
\label{section:rw-renderedcontent}

Several methods have been proposed to improve the visual fidelity of rendered content or to upsample a rendering performed at a lower resolution. Weiss \emph{et al.}\cite{Weiss19} used a deep learning-based architecture to upscale the resolution for isosurface rendering. Nvidia recently introduced a super-sampling technique that uses a deep neural network and temporal history to accumulate samples \cite{DLSS}. Similarly, Xiao \emph{et al.}\cite{Metapaper} demonstrated up to 4x4 upsampling of highly aliased input. These methods, however, perform image reconstruction for surface-only rendered content. In our work, we focus on performing up to 8x8 super-resolution of volumetric visualization with high visual and temporal fidelity. Furthermore, most of these above methods propose to use motion information between frames to use temporal history, however, computing screen space motion information for volumetric rendering is not straightforward.

\section{Methodology}
\label{section:methodology}

In this section, we describe the overall framework of our system.

\subsection{ Direct Volume Rendering Framework}
\label{section:meth-dvr}

In our DVR framework, we cast rays from the camera through pixels of the view-port. When the ray reaches the volume contained in an axis-aligned bounding box, the ray is sampled via ray marching, i.e., stepped along at equal distances. At every step of the ray, a transfer function maps the interpolated intensity value at that position to an RGBA vector. As the ray steps through the volume, a local gradient is combined with a local illumination model to provide realistic shading of the object. The final pixel value is computed using front-to-back compositing of the acquired color and alpha (opacity) values along the ray. The ray is terminated early if either the accumulated opacity reaches close to 1 or the ray leaves the volume.

The issue with high-quality super-resolution for rendered content is that the information at the to-be-interpolated pixels at the target resolution is completely missing and since pixels are point-sampled, they are extremely aliased at geometry edges, especially at low resolution. An effective way to handle these aliasing artifacts is temporal anti-aliasing (TAA) which attempts to gather multiple samples per pixel by distributing the computations across multiple frames. Motivated by this, we implement a similar technique to perform super-resolution i.e., compute and gather multiple sub-pixel samples across frames and feed this information to our super-resolution network to upscale the low-resolution rendering. However, for volumetric rendering, accumulating samples from previous frames presents a few challenges which we discuss in the sections below.

\subsubsection{Motion Vector and Depth} 
\label{section:meth-motionanddepth}

In rendering, a motion vector defines an analytically computed screen-space location where a 3D point that is visible at the current frame $i$ would appear in the previous frame $i-1$. The main principle of temporal methods to perform either anti-aliasing or upsampling is to compute multiple sub-pixel samples across frames, and then combine those together for the current frame. The samples from the previous frame are reprojected using the motion vector to the current frame. The input to our renderer is static volumetric data without any motion of its own, so performing reprojection using the motion vector depends entirely on the camera transformation matrices and depth information. Unfortunately for direct volume rendering, we lack this depth information since we are not looking at a single position in the world space but a number of points in the volume along the ray. Hence, computing motion vectors to perform reprojection is challenging. 

\begin{figure*}[t]
  \centering
  \includegraphics[width=0.5\linewidth]{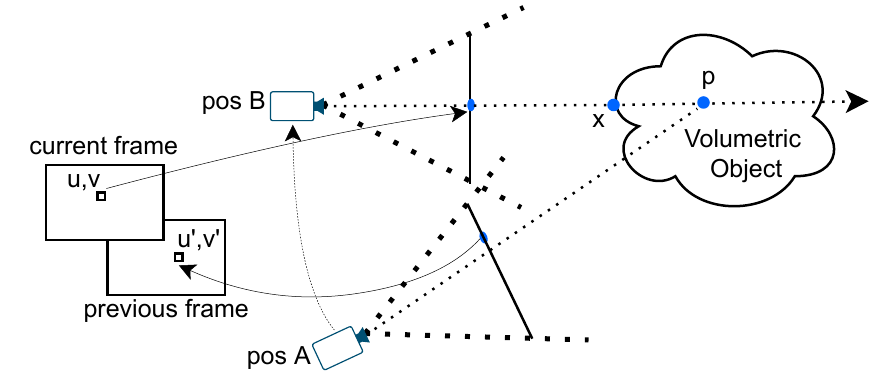}
  \centering
  \caption{Camera movement around a volumetric object from $pos A$ to $pos B$. Point $x$ is the first hit point on the volumetric object when the ray passes through the volume for the current camera position, while $p$ is a point inside the volume where the alpha value is maximum along the ray. }
  \label{fig:motionanddepth}

\end{figure*}

To overcome this, we implement a naive approach where we use the point of maximum alpha along the ray to perform reprojection. Since this position will have the maximum contribution to the final accumulated sample, we found that this quasi-depth information computed using this heuristic gives an acceptable approximation for the estimation of motion vector. We start from the current frame coordinate $u,v$  as shown in figure \ref{fig:motionanddepth}. Once we have the world space position for the point $p$ in space where we have maximum alpha along the ray, we can use previous camera transformation matrices to reproject this position back to previous frame coordinates $u',v'$. The difference between the two frame coordinates gives us the screen space motion vector due to camera movement. 

\subsubsection{Disocclusion and Ghosting} 
\label{section:meth-ghosting}

Once we have the motion vector between two consecutive frames, we additionally incorporate temporal anti-aliasing to our final rendering with an additional compute shader call, thus adding a post-processing pass to our DVR pipeline. We utilize the history color buffer and motion vector to gather samples from the previous frame and combine them with the samples in the current frame. History samples can sometimes be invalid. Trivially accepting all of the history samples causes ghosting artifacts in the final rendered image because of disocclusion. As we move the camera, regions of the volume that were not previously visible may come into view. To address this issue, similar to \cite{Lottes11}\cite{Malan12}, we resort to using neighborhood clamping which makes the assumption that colors within the neighborhood of the current sample are valid contributions to the accumulation process. Specifically, we implemented 3*3 neighborhood clamping which produced reasonably effective results for our volume rendering case. 

\begin{figure*}[t]
  \centering
  \includegraphics[width=0.8\linewidth]{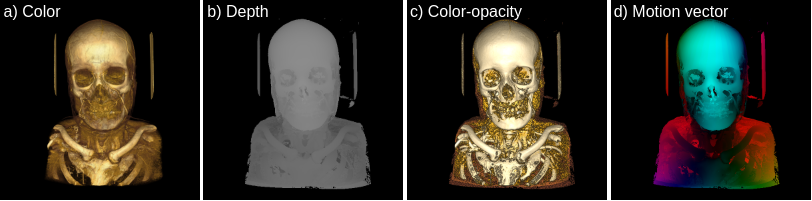}
  \centering
  \caption{Input feature images from the training dataset. From left to right: a) Final rendered color image with 3 channels RGB, b) Depth with a single channel and c) color-opacity vector (opacity is not shown) with 4 channels RGBA at the position where alpha is maximum along the ray; d) Example Motion vector image with 2 channels. 
  }
  \label{fig:input_features}

\end{figure*}

\subsubsection{Supplementary Features} 
\label{section:meth-supplementaryfeatures}

Previous works on reconstruction networks for surface data \cite{Metapaper}\cite{Koskela19}\ have shown that supplementing a network with additional features improves the overall performance of the network. This motivates us to opt for a few supplementary features adapted to our volumetric case. Xiao \emph{et al.}\cite{Metapaper} showed that the reconstruction network benefits with depth as an additional input to the network, but as discussed in section \ref{section:meth-motionanddepth},  depth information is not well defined for the volumetric case, so we resort to using the depth at the point of maximum alpha value along the ray since the final rendering will have more contribution from this point. Additionally, we also save color and opacity values at this point. When adding them as input, we are able to obtain additional gains with our network (section \ref{section:eval-additionalinput}). 

In addition to feeding the network with rendered frames and supplementary features from the current and the previous time steps, we also provide a screen space 2D motion vector which is used to warp the previous frames to the current frame. Using optical flow or motion estimation is common in the video super-resolution literature (section \ref{section:relatedWork}) to capture the temporal dependency between successive frames and to reduce the complexity of the network. Figure \ref{fig:input_features} shows all types of inputs that our network receives.

\begin{figure*} [t]

    \centering
        \begin{minipage}{.64\textwidth}
            \centering
            \includegraphics[width=1\linewidth]{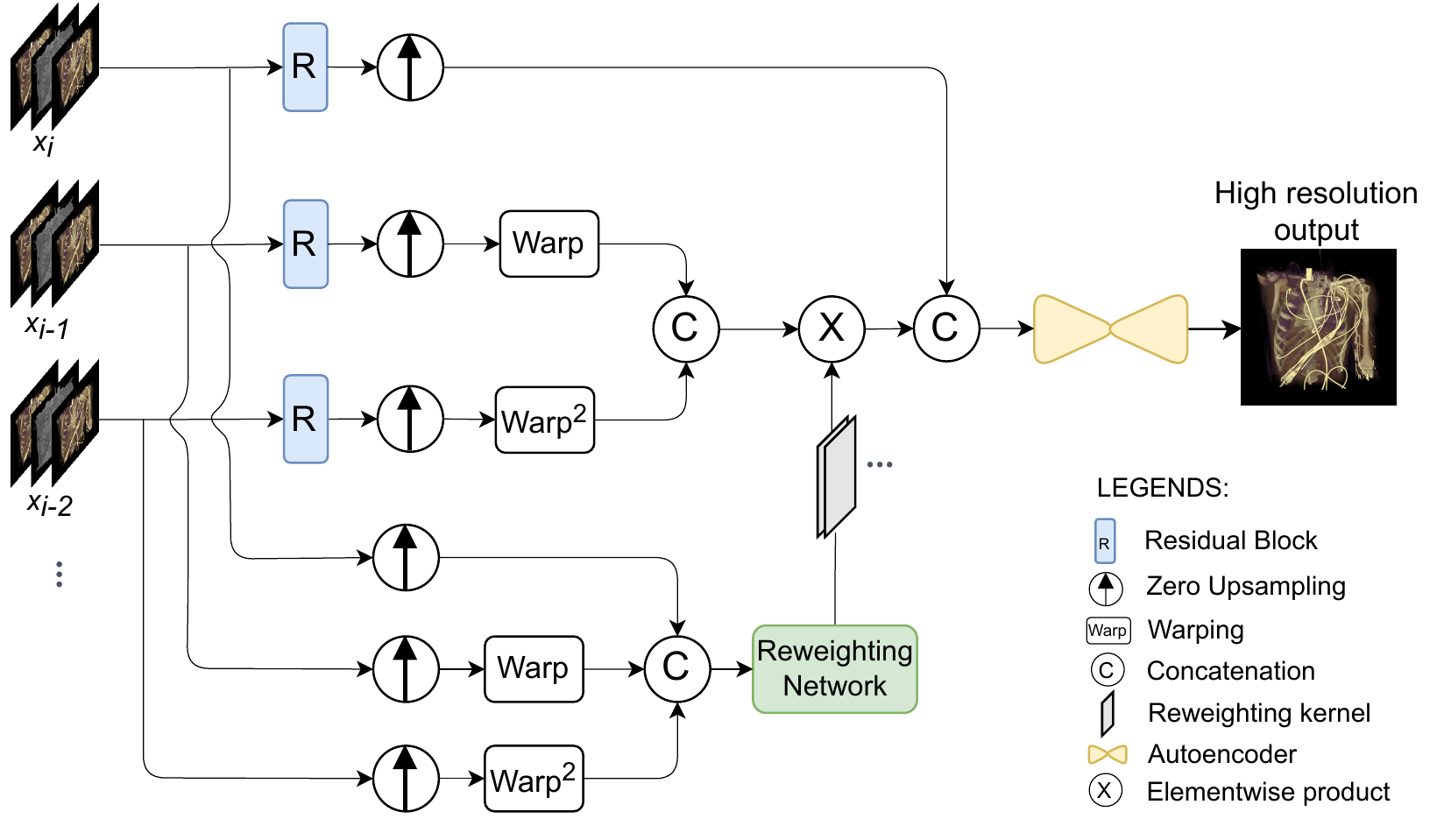}
            \centering
            \caption{Overall network architecture with components inspired from 
            Xiao \emph{et al.}\cite{Metapaper} and Hofmann \emph{et al.}\cite{Hofmann20} 
            }
            \label{fig:overallnetwork}
        \end{minipage}\qquad
        \begin{minipage}{.30\textwidth}
            \centering
            \includegraphics[width=0.9\linewidth]{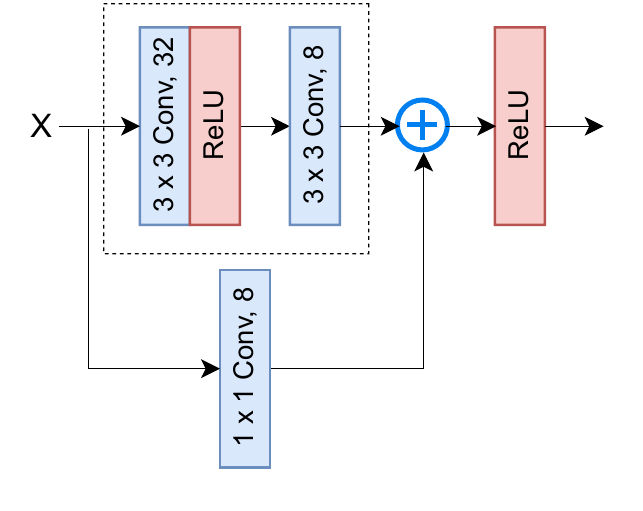}
            \caption{Residual Block}
            \label{fig:residualblock}
        \end{minipage}

    \centering
        \begin{minipage}{.40\textwidth}
            \centering
            \includegraphics[width=0.9\linewidth]{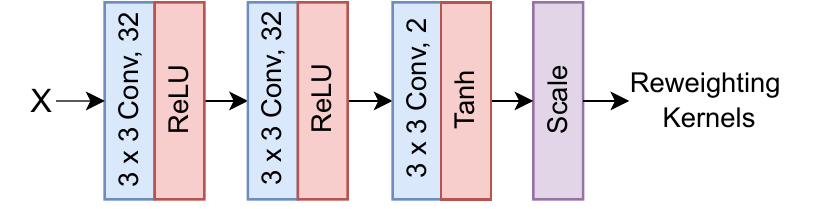}
            \caption{Reweighting Network}
            \label{fig:reweightingnetwork}
        \end{minipage}\qquad
        \begin{minipage}{.54\textwidth}
            \centering
            \includegraphics[width=1\linewidth]{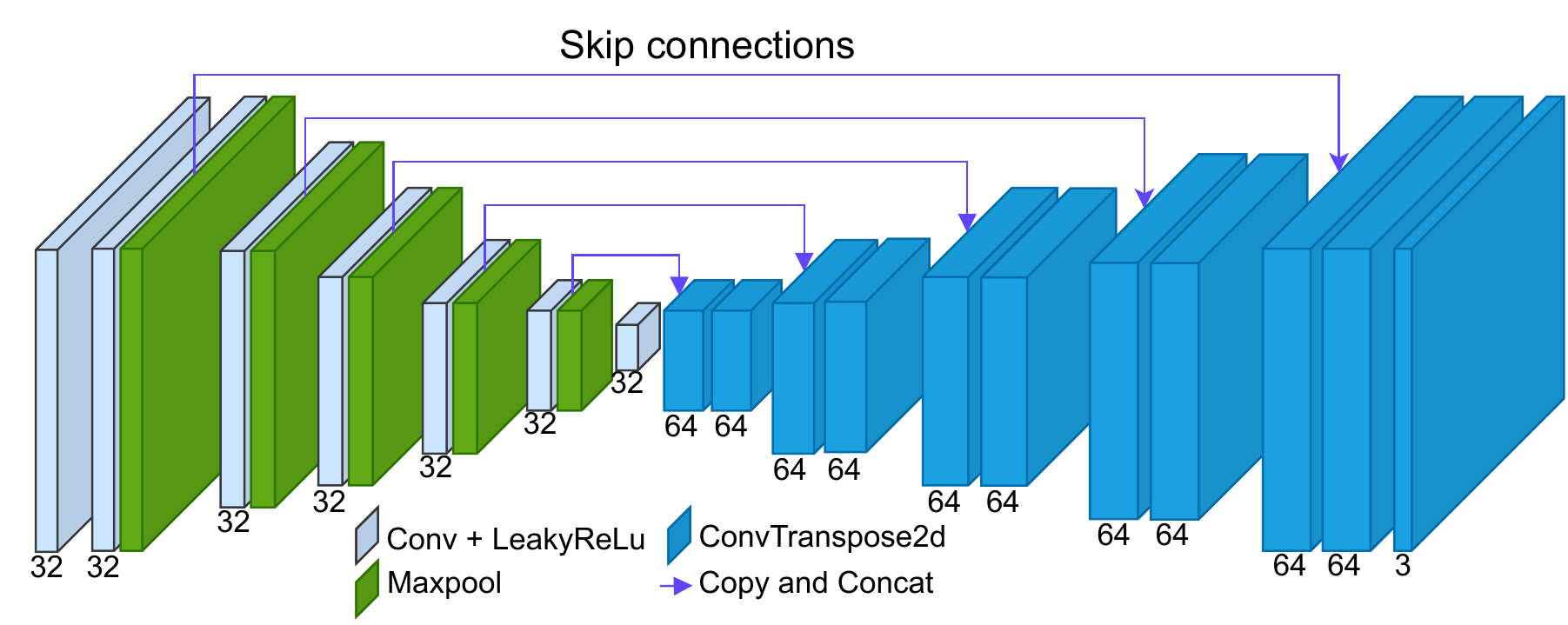}
            \centering
            \caption{Autoencoder for reconstruction of high resolution image }
            \label{fig:autoencoder}
        \end{minipage}

\end{figure*}

\subsection{ Network Architecture}
\label{section:meth-networkarch}

Figure \ref{fig:overallnetwork} depicts the data flow through our network. The overall network architecture has been inspired from Xiao \emph{et al.}\cite{Metapaper} with a number of modifications to suit our needs. We implement residual blocks to extract features from the input since they are easier to train and allow a better flow of information due to the presence of shortcut connections \cite{He16}. For our reconstruction network, we adopt a similar autoencoder architecture by Hofmann \emph{et al.} \cite{Hofmann20} which has been successfully applied to volumetric data. For the loss formulation, we implement Charbonnier loss because of its benefits mentioned in section \ref{section:meth-loss}.

\subsubsection{Residual Block} 
\label{section:meth-residualblock}

The first component of the network is a residual block which is used to extract features from the input frames, where by `input frames', we mean all the rendered color images from current and previous frames with their supplementary features excluding motion vector. The residual block we use in our network has two 3x3 convolutional layers. Each convolutional layer is followed by a rectified linear unit (ReLU) activation function. After the second convolutional layer, the output from the layer is added together with the input to the residual block, before sending it to the final ReLU activation function. To transform the input into the desired shape for the addition operation, we introduce an additional 1x1 convolutional layer in the skip connection.

\subsubsection{Zero Upsample and Warping}
\label{section:meth-zeroupsample}
We implement zero upsampling technique \cite{Metapaper} to upscale the low-resolution input to the target resolution. In zero upsampling, every pixel in the low-resolution space is upsampled to be surrounded by pixels with zero values in high-resolution space. Once all the input frames and the feature maps (extracted from the residual block) are upsampled to target resolution, the previous frames and the corresponding feature maps are processed further with the warping module, where they are backward warped to align with the current frame with the help of motion vectors. All input frames (after zero upsampling and warping) are then concatenated and fed to a reweighting network as shown in figure \ref{fig:overallnetwork}.

\subsubsection{Reweighting Network}
\label{section:meth-reweightingnetwork}
 
As discussed in section \ref{section:meth-ghosting}, there are a few limitations associated with using motion vectors that prevent its direct use for accumulating history samples. In addition to disocclusion and ghosting, motion vectors do not reflect shading and lighting changes between two frames. To address these issues, we leverage a recent work in neural upsampling \cite{Metapaper} which uses a reweighting network to weed out the inconsistent samples. The reweighting network is shown in figure \ref{fig:reweightingnetwork}. It is a 3 layer convolutional network that generates a pixel-wise reweighting channel for each previous frame. For example, for two previous frames used in our network, we obtain two reweighting channels from the reweighting network. Each of these reweighting channels undergoes elementwise multiplication with all the channels of each of the previous frame’s feature maps (after zero upsampling and warping). The result is concatenated with the current frame’s feature map and fed as an input to an autoencoder.

\subsubsection{Autoencoder}
\label{section:meth-autoencoder}
For the reconstruction of high-resolution images using the concatenated result from section \ref{section:meth-reweightingnetwork}, we adopt a similar autoencoder network from Hofmann \emph{et al.} \cite{Hofmann20}. It uses a fully convolutional encoder and decoder hierarchy with skip connections as shown in figure \ref{fig:autoencoder}.

\subsubsection{Loss Function}
\label{section:meth-loss}
We use Charbonnier loss \cite{Charbonnier} to quantify the error between the high-resolution output and the given ground truth image. Charbonnier loss is known to be insensitive to outliers and for super-resolution tasks, experimental evaluation has shown that it provides better PSNR/SSIM accuracies over other conventional loss functions \cite{Anagun19}.

\begin{equation}\label{eq:loss}
L = \frac{1}{N} \sum_{i=0}^{N} \rho(y_i - z_i) ,
\end{equation}

where, $\rho(x) = \sqrt{x^2 + \epsilon^2}, \epsilon=1\times10^{-8} , z_i $
denotes the ground truth high resolution frame, 
and $N$ denotes the number of pixels.

\section{Dataset}
\label{section:dataset}

In order to generate a high quality dataset, we incorporate 3 different volumetric data (CTA-Cardio: 512x512x321, Manix: 512x512x460, CTA Abdomen Panoramix: 441x321x215) with different transfer functions. We render 36 videos from each volume data and each video contains 100 frames. Each of these videos start from a random camera position in the scene that is selected from a large candidate pool. We split the dataset generated from each scene into 3 sets: training $(80\%)$, validation $(10\%)$, and test $(10\%)$.


For ground truth high-resolution images, we render the volume data at 1920x1920 resolution with temporal anti-aliasing turned on. For low-resolution input, the temporal anti-aliasing feature is turned off and the images are rendered at varying resolutions: 480x480, 240x240, and 120x120. In image and video super-resolution literature, it is common practice to use blurred and downscaled versions of the original high-resolution image as low-resolution input to the network. In contrast, our low-resolution input is directly generated from our volume renderer. We train different networks to perform 4x4, 8x8, and up to 16x16 super-resolution with the respective combination of low and high-resolution images.

\begin{table}[t]
\small
\centering
\caption{Quantitative comparison between two networks: with and without the use of additional RGBA information from the point of maximum alpha }
\label{table:additionalinput}
\begin{tabular}{|l|l|l|l|l|}
\hline
& \multicolumn{2}{|l|}{With additional information}
  &  \multicolumn{2}{|l|}{Without additional information} \\
\hline
Volume Dataset          &   PSNR(dB)    &   SSIM    &   PSNR(dB)    &   SSIM    \\
\hline
CTA-Cardio              &   38.09       &   0.9705  &   37.07       &   0.9683  \\
Manix                   &   37.92       &   0.9651  &   36.96       &   0.9631  \\
CTA-Abdomen             &   31.89       &   0.9560  &   31.44       &   0.9557  \\
\hline
\end{tabular}
\end{table}

\begin{figure*}[t]
  \centering
  \includegraphics[width=0.7\linewidth]{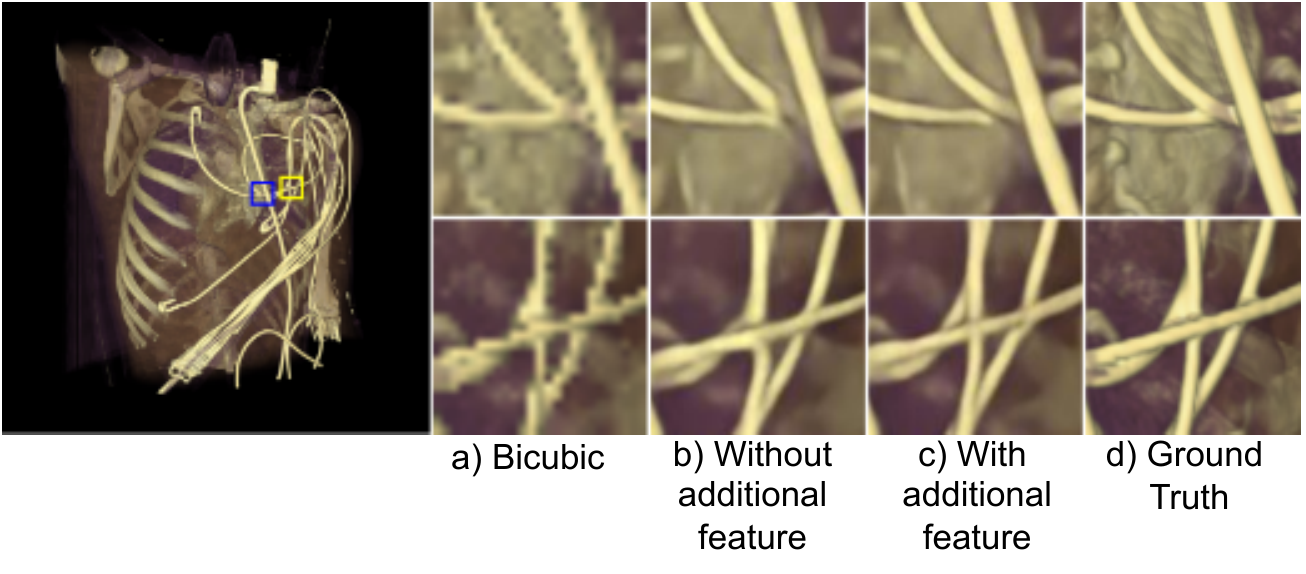}
  \centering
  \caption{Visual comparison for 8x8 upscaling with different techniques on the CTA-Cardio dataset. Images on the top and bottom row (enlarged sections of the blue and yellow boxes respectively) are from two different sections of CTA-Chest. a) represents input upscaled with bicubic interpolation. Comparing b) and c), we notice improved edges and details in the upscaled image when the super-resolution network is supplemented with additional RGBA information from the point of highest contribution  }
  \label{fig:additionalinput}
\end{figure*}

\section{Evaluation}
\label{section:evaluation}

For the evaluation, we compare the performance of different variants of our network on Peak Signal To Noise Ratio (PSNR) and Structural Similarity Index (SSIM). The reported results are observed on the validation set.

\subsection{Performance Gain with Additional Feature at the Input}
\label{section:eval-additionalinput}

As discussed in section \ref{section:meth-supplementaryfeatures}, including auxiliary features at the input generally benefits the network to achieve additional performance gain. In table \ref{table:additionalinput}, we compare the observed performance metrics for all the three datasets when we include an additional feature at the input. The additional feature is the RGBA information obtained from the point of highest contribution along the ray. In addition to quantitative improvement in both PSNR and SSIM, we also observe improved edges and details in the reconstructed images as shown in figure \ref{fig:additionalinput}.

\begin{table}[t]
    \centering
    \caption{Performance gain achieved with additional previous frames on CTA-Abdomen(table on the left) and CTA-cardio (table on the right) Dataset for 4x4 upsampling. $N$ denotes the number of previous frames.
    \label{table:additionalframe}
    }
        \begin{tabular}{|l|l|l|l|}
        \hline
        $N$  &   1       &   2       &   3       \\
        \hline
        PSNR (dB)               &   31.86   &   32.60   &   32.98   \\
        SSIM                    &   0.9552  &   0.9606  &   0.9638  \\
        \hline
        \end{tabular}
    \quad
        \begin{tabular}{|l|l|l|l|}
        \hline
        $N$  &   1       &   2       &   3       \\
        \hline
        PSNR (dB)               &   39.35   &   39.94   &   40.49   \\
        SSIM                    &   0.9690  &   0.9755  &   0.9783  \\
        \hline
        \end{tabular}
\end{table}

\begin{figure*}[t]
  \centering
  \includegraphics[width=0.7\linewidth]{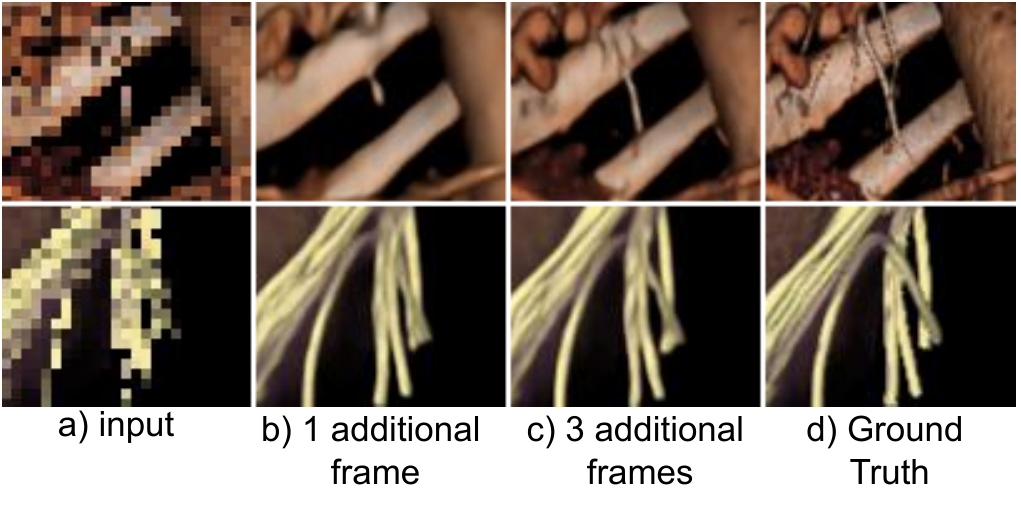}
  \centering
  \caption{Visual comparison for 4x4 upsampling on CTA-Abdomen and CTA-Cardio. a) is the input to two different networks: one takes a single previous frame whose output is in b), and the other takes up to 3 previous frames whose output is in c).  }
  \label{fig:additionalframe}
\end{figure*}

\subsection{Performance Gain with Additional Previous Frames}
\label{section:eval-additionalframe}

In table \ref{table:additionalframe}, we report the quantitative evaluation of three different networks, each of which takes a different number of previous frames. We are able to make additional gains on both PSNR and SSIM with additional previous frames supplied to the network. In addition to improvements in the quality of the reconstructed image (figure \ref{fig:additionalframe}), incorporating additional frames also improved the temporal stability of the reconstructed video sequence
(video: \href{https://youtu.be/1FZCQG0SBac}{youtu.be/1FZCQG0SBac}).

\begin{table}[t]
\centering
\caption{Quantitative comparison for various upsampling ratios on the Manix Dataset }
\label{table:upsamplingratio}
\begin{tabular}{|l|l|l|l|}
\hline
Upsampling Ratio    &   4x4     &   8x8     &   16x16   \\
\hline
PSNR(dB)            &   42.37   &   37.92   &   33.65   \\
SSIM                &   0.9787  &   0.9651  &   0.9471  \\
\hline
\end{tabular}
\end{table}

\begin{figure*}[t]
  \centering
  \includegraphics[width=0.7\linewidth]{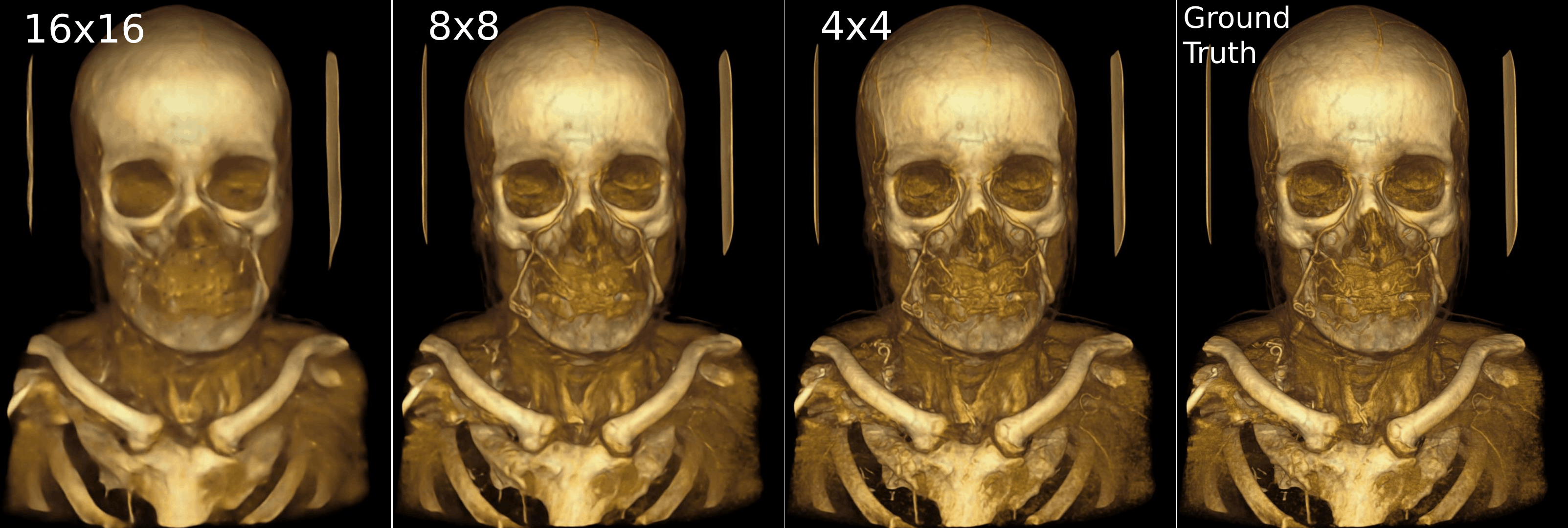}
  \centering
  \caption{Visual comparison for various upscaling ratios on the Manix dataset. For all images, target resolution was 1920x1920. }
  \label{fig:upscalingratio}

\end{figure*}

\subsection{Upsampling Ratio}
\label{section:eval-upsamplingratio}

To test the limits of our super-resolution network, we take it one step further and perform up to 16x super-resolution. The observed PSNR and SSIM metrics are shown in table \ref{table:upsamplingratio}. The target resolution for all the upsampling ratios was the same 1920x1920, while the input resolution varied according to the upsampling ratio. As the upsampling ratio increases, the quality of the reconstructed images steadily deteriorates and the network is unable to reconstruct the low-level features which are also evident from the images shown in figure \ref{fig:upscalingratio}.

\section{Conclusion and Future Work}
\label{section:conclusion}

In our work, we introduced a new pipeline to perform super-resolution for medical volume visualization. Our approach includes several adjustments tailored to the volumetric nature of the data. Despite our improvements, there are numerous future works that could be performed from here. Currently, all of our volumetric datasets are static volumetric data without any motion of their own. The introduction of dynamic volume will add more challenges to the system. Another future extension could be supplementing our network with additional volumetric features from multiple depths inside the volume. We believe this can further improve the reconstruction ability of the super-resolution network.  

Furthermore, it should be noted that our system was designed for offline application and less importance was given to run-time performance. The current implementation of our network is able to perform super-resolution at an interactive frame rate of ~10 fps (0.1018 seconds per frame). With run-time optimizations and integration of TensorRT, which can provide up to 6x faster accelerated inference\cite{Tensorrt}, our system has the potential to achieve real-time frame-rate.

\bibliographystyle{splncs04}
\bibliography{references}

\end{document}